# Computational strategies for dissecting the high-dimensional complexity of adaptive immune repertoires

Enkelejda Miho[a], Alexander Yermanos[a], Cédric R. Weber[a], Christoph T. Berger[b,c], Sai T. Reddy[a,#], Victor Greiff[a,d,#]

[a] ETH Zürich, Department for Biosystems Science and Engineering, Basel, Switzerland
[b] Department of Biomedicine, University Hospital Basel, Basel, Switzerland
[c] Clinical Immunology, Department of Internal Medicine, University Hospital Basel, Basel, Switzerland
[d] Department of Immunology, University of Oslo, Oslo, Norway
[#] Correspondence: victor.greiff@medisin.uio.no, sai.reddy@ethz.ch

**Abstract**

The adaptive immune system recognizes antigens via an immense array of antigen-binding antibodies and T-cell receptors, the immune repertoire. The interrogation of immune repertoires is of high relevance for understanding the adaptive immune response in disease and infection (e.g., autoimmunity, cancer, HIV). Adaptive immune receptor repertoire sequencing (AIRR-seq) has driven the quantitative and molecular-level profiling of immune repertoires thereby revealing the high-dimensional complexity of the immune receptor sequence landscape. Several methods for the computational and statistical analysis of large-scale AIRR-seq data have been developed to resolve immune repertoire complexity in order to understand the dynamics of adaptive immunity. Here, we review the current research on (i) diversity, (ii) clustering and network, (iii) phylogenetic and (iv) machine learning methods applied to dissect, quantify and compare the architecture, evolution, and specificity of immune repertoires. We summarize outstanding questions in computational immunology and propose future directions for systems immunology towards coupling AIRR-seq with the computational discovery of immunotherapeutics, vaccines, and immunodiagnostics.

**Introduction**

The adaptive immune system is responsible for the specific recognition and elimination of antigens originating from infection and disease. Molecular recognition of antigens is achieved through the vast diversity of antibody (B-cell receptor, BCR) and T-cell receptors (TCR). The genetic diversity of these adaptive immune receptors is generated through a somatic recombination process that acts on their constituent V, D, and J segments (1,2). During the gene rearrangement process, additional sequence diversity is created by nucleotide deletion and addition, resulting in a potential diversity of >$10^{13}$ unique B- and T-cell immune receptor sequences (3–6). The adaptive immune repertoire often refers to the collection of all antibody and T-cell immune receptors within an individual, and represents both the ongoing and past immune status of an individual. Current threats, for example of pathogenic nature, are countered by B- and T-cell clonal expansion and selection (7), whereas past ones are archived in immunological memory compartments (8). Immune repertoires are highly dynamic. They are constantly evolving within the repertoire sequence space, which is defined as the set of all biologically achievable immune receptor sequences. Repertoire dynamics and evolution span several orders of magnitude in size (germline gene to clonal diversity), physical components (molecular to cellular dynamics) and time (short-lived responses to immunological memory that can persist for decades (9–14)).

The quantitative resolution of immune repertoires has been fueled by the advent of high-throughput sequencing (2,15–20). Since 2009, high-throughput adaptive immune receptor repertoire sequencing (AIRR-seq) has provided unprecedented molecular insight into the complexity of adaptive immunity by generating datasets of 100 millions to billions of reads (6,21,22). The exponential rise in immune repertoire data has correspondingly led to a large increase in the number of computational methods directed at dissecting repertoire complexity (Figures 1,2) (23). Immune repertoire sequencing has catalyzed the field of computational and systems immunology in the same way that genomics and transcriptomics has for systems and computational biology (23). To date, the computational methods that have been developed and applied to immune repertoires relate to (i) the underlying mechanisms of diversity generation, (ii) repertoire architecture, (iii) antibody evolution, and (iv) molecular convergence.

This review provides an overview of the computational methods that are currently being used to dissect the high-dimensional complexity of immune repertoires. We will treat only those methods that are down-stream of data preprocessing (although currently there is no consensus on standard operating pre-processing procedures, please refer to recent reviews on these subjects (2,17,24)). Specifically, this review centers on computational, mathematical and statistical approaches used to analyze, measure and predict immune repertoire complexity. The description of these methods will be embedded within the main areas of immune repertoire research. Given that the genetic structure of antibody and T-cell receptors is very similar, the majority of the methods illustrated in this review can be applied both in the context of antibody and T-cell studies. Exceptions to this rule are clearly stated.

**Measuring immune repertoire diversity**

The immense diversity is one of the key features of immune repertoires and enables broad antigen recognition breadth (Figure 1A, 2A). The maximum theoretical amino acid diversity of immune repertoires is $\approx 10^{140}$ (calculated as $20^{110}*2$). The calculation takes into account the 20 unique amino acids, the 110 amino acids long variable region of immune receptors, and the 2 variable regions composing each receptor ($IGV_L$-$IGV_H$ or $TCRV\alpha$-$TCRV\beta$) (25). This enormous diversity, however, is restricted in humans and mice by a starting set of V, D, J gene segments leading to a potential diversity of about $10^{13}$–$10^{18}$ (3–6,26–30). Only a fraction of the potential diversity is represented at any point in time in any given individual: the number of B and T cells is restricted (human: $10^{11-12}$) and the number of different clones, depending on clone definitions, reaches about $10^9$ in humans and $10^{6-7}$ in mice (3,5,6,31). The study of immune repertoire diversity ranges from the study of (i) the diversity of the building blocks of immune repertoires (V, D, J segments) and antibody lineage reconstruction (ii) to the mathematical modeling of VDJ recombination (iii) to the estimation of the theoretical and biologically available repertoire frequency diversity (32). Together, these subfields of repertoire diversity analysis have expanded our analytical and quantitative insight into the creation of naïve and antigen-driven antigen receptor diversity.

Accurate quantification of repertoire diversity relies first and foremost on the correct annotation of sequencing reads. Read annotation encompasses multiple steps: (i) calling of V, D, J segments, (ii) subdivision into framework (FR) and complementarity determining regions (CDR), (iii) identification of inserted and deleted nucleotides in the junction region, (iv) the quantification of the extent of somatic hypermutation (for antibodies). VDJ annotation tools were recently reviewed by Greiff et al. and Yaari et al. (17,24). An updated version is currently maintained on the B-T.CR forum (http://b-t.cr/t/list-of-v-d-j-annotation-software/18). The B-T.CR forum is an AIRR-seq community platform for community-edited Wiki pages related to datasets and analysis tools as well as scientific exchange on current relevant topics in AIRR-seq (33,34).

Accurate antigen receptor germline gene genotyping is crucial for predicting adaptive immunity (personalized and precision medicine) in the genetically diverse human population (30,35). All VDJ annotation tools rely, at least partly, on a reference database of germline gene alleles. A reference database that is not identical to that of the individual from which the sequencing data is being annotated bears the potential of inaccurate annotation. This could affect for example the accuracy of the calling of V, D, and J genes and alleles as well as the quantification of somatic hypermutation. Antibody gene allele variation has also been linked to differential effectiveness of the humoral immune response (30,35). Indeed, an increasing number of human germline gene alleles – representing one or several single nucleotide polymorphisms – has been recently detected (30,36–39). These discoveries call into question the widely adopted practice of using one central germline reference database containing a more or less static set of non-personalized germline gene alleles. To address this problem, Corcoran et al. developed a software package (IgDiscover), which employs a cluster identification approach to reconstruct *de novo* from an AIRR-seq dataset the corresponding V-gene germline database – all without *a priori* knowledge of existing germline gene databases (40). By doing so, they detected extensive individual germline gene differences among rhesus macaques (40). Complementarily, Gadala-Maria et al. developed TiGER (Tool for Ig Genotype Elucidation via Rep-Seq), which detects novel alleles based on mutation pattern analysis (37). In contrast to IgDiscover, TiGER uses initial VDJ allele assignments with existing databases and software. Extending the analysis of germline gene diversity to the population level, Yu et al. built Lym1K, which is a database that combines validated alleles with novel alleles found in the 1000 Genomes Project (41,42). In addition to database-centered approaches, probabilistic annotation enabled the detection of novel IgV-genes and led to the discovery that substitution and mutation processes are (although reproducible across individuals) segment and allele-dependent, thus further refining VDJ annotation and downstream diversity measurement (4,43–46).

As a direct application in fundamental immunology, the advent of AIRR-seq has enabled the opportunity to quantitatively test whether VDJ recombination is a random process. The ability to generate large datasets allowed several studies to show evidence of biases in VDJ recombination, as some germline gene frequencies (and combinations thereof) were found to occur more often than others (6,21,47–49). In order to mathematically model the process of VDJ recombination in both T and B cells, Elhanati et al. and Murugan et al. have employed techniques borrowed from statistical physics (maximum entropy, Hidden Markov and probabilistic models) (4,5,44) to uncover the amount of diversity information inherent to each part of antibody and TCR sequences (entropy decomposition). VDJ recombination probability inference was mostly performed on non-productive sequences (e.g., out-of-frame, containing stop codon) as these receptors were assumed to be exempt from selection, thus representing unselected products of the generation process (4).

The deep sequence coverage of AIRR-seq has also led to the discovery of public clones or clonotypes – sequences that are shared across two or more individuals (6,50–52). The existence of naïve and antigen-associated public clones signifies a predetermined reduction in *a priori* genetic and antigen-driven immune receptor diversity (6). Although the exact definition of what constitutes a "public clone" is debatable (53), advancements have been made in understanding the generation and structure of public B- and T-cell clonotypes. By quantifying VDJ recombination probabilities as described above, Elhanati et al. have suggested that the emergence of public clonotypes is a direct consequence of the underlying VDJ recombination bias (54). The inference of VDJ recombination statistics of naïve B- and T-cell populations may be of use in vaccination studies for helping distinguish public *antigen-specific* clonotypes from *genetically (naïve) predetermined* ones. If feasible, such an approach might render the need of a healthy control cohort for determining *naïve* public clones superfluous (46,55). Complementarily, Greiff et al. have demonstrated extensive VDJ recombination bias by support vector machine analysis. Specifically, it showed that both public and private clones possess predetermined sequence signatures independent of mouse strain, species and immune receptor type (antibody, TCR). These sequence signatures were found in both naïve and antigen-selected B-cell compartments, which might suggest that naïve recombination bias exerts a stronger diversity-constricting effect than antigen-driven evolution (56).

While the above-described methods of immune repertoire diversity analysis are relatively new, the quantification and comparison of clonotype diversity has been already studied in the era preceding high-throughput sequencing platforms by borrowing and adapting from mathematical ecology (57–60). The first step to quantifying clonal repertoire diversity is the definition of clonotype. Definitions of clonotype used in the literature range from the exact amino acid CDR3 to clusters of (e.g., CDR3) sequences to the sequence of entire variable chain regions ($IGV_L$-$IGV_H$ or $TCRV\alpha$-

TCRVβ) using methods ranging from likelihood-based lineage inference to distance-based measures. A complete list of clonotyping tools has been compiled on the B-T.CR forum (http://b-t.cr/t/list-of-b-cell-clonal-identification-software/22). The debate on what constitutes a clonotype is ongoing and beyond the scope of this review. The interested reader is kindly referred to two extensive reviews (17,61) and a recent report by Nouri and Kleinstein, who have developed a flexible user-defined method for clonotype identification (62).

To measure clonotype diversity, diversity indices are used (detailed reviews on diversity indices have been recently published (17,24)). Briefly, diversity indices enable the comparison of repertoire diversity by parameterizing the repertoire space. They thus overcome the problem of mostly clonally distinct repertoires (63). Several dedicated software packages exist for diversity index calculations (64–67). Briefly, the Diversity ($^\alpha D$) of a repertoire of $S$ clones is usually calculated as follows: $^\alpha D = \left(\sum_{i=1}^{S} f_i^\alpha\right)^{\frac{1}{1-\alpha}}$ (Hill-Diversity), where $f_i$ is the frequency of the $i$th clone weighted by the parameter $\alpha$. Special cases of this Diversity function correspond to popular diversity indices in the immune repertoire field: species richness ($\alpha = 0$), and the exponential Shannon-Weiner ($\alpha \to 1$), inverse Simpson ($\alpha = 2$), and Berger-Parker index ($\alpha \to \infty$). The higher the value of alpha, the higher becomes the influence of the more abundant clones on the $^\alpha D$. Due to the mathematical properties of the Diversity function (Schur concavity, (68)), different repertoires may yield *qualitatively* different $^\alpha D$ values depending on the Diversity index used (Figure 1 in Greiff et al., (63)). Thus, for any discriminatory diversity comparisons, at least two Diversity indices should be considered. Diversity *profiles*, which are collections (vectors) of several Diversity indices, have been suggested to be superior to *single* diversity indices, when comparing clonal diversity (63,64,69). Using hierarchical clustering, $\alpha$-parametrized diversity profiles have been shown to faithfully capture the shape of a repertoire's underlying clonal frequency distribution, which represents the state of clonal expansion (63). Thus, diversity profiles can serve as a parameterized proxy for a repertoire's state of clonal expansion. Additionally, Mora and Walczak showed that the Rényi entropy (the mathematical foundation of Hill-Diversity profiles) can be constructed, in some cases, from rank-frequency plots (70) thereby establishing a direct mathematical link between clonal frequency distribution and diversity indices. Another interesting novel diversity analysis method is the *clonal plane* and the *polyclonal monoclonal diversity* index developed by Afzal et al. (71). Briefly, these two related concepts represent repertoire diversity in a coordinate system spanned by species richness and evenness. This allows a visually straightforward identification of polyclonal and oligoclonal samples.

Although clonal frequency distributions can, in most cases, not be compared directly across individuals due to restricted clonal overlap, their mathematical description has been the object of several studies. Specifically, clonal frequency distributions were found to be power-law distributed, with a few abundant clones, and a large number of lowly abundant clones (63,72–74). Furthermore, Schwab et al. showed analytically and via numerical simulations that Zipf-like distributions, a subclass of power-law distributions, arise naturally if fluctuating unobserved variables affect the system (e.g., a variable external antigen environment influencing the observed antibody repertoire) (75). Indeed, it could be shown that clonotype diversity (or state of clonal expansion) contains antigen-associated information on the host immune status (6,63,76).

In order to compare differences between diversity profiles, one should also consider resampling strategies as implemented the R package *Change-O* by Gupta et al. These allow the determination of confidence areas around each diversity profile (64,77) in the presence of differently sized repertoires. The precision of diversity calculation in case of incomplete sampling is of special importance when gaining information on human repertoires, which often are restricted to the isolation of a limited number of B and T cells from peripheral blood (17,78–80).

Although the quantification of diversity is one of the more mature subfields of computational repertoire immunology, numerous open questions remain: (i) Diversity has been measured from many different perspectives (germline gene diversity, state of clonal expansion, clonal size), thus capturing different dimensions of the repertoire diversity space. Is it possible to devise a universal metric that synthetizing different aspects of immune repertoire diversity into one? Such a metric would be very useful for repertoire-based immunodiagnostics. (ii) Hidden Markov and Bayesian (probabilistic) approaches have been used for modeling VDJ recombination. Those approaches, however, capture only short-range sequence interactions. Might, therefore, recurrent neural network approaches be more appropriate to model the immune repertoire sequence space given their ability to account for sequence interactions of arbitrary length (81,82)? (iii) Finally, we still have only very superficial insight into the biological diversity of antigen-specific repertoires and the combination rules of $IGV_L/IGV_H$ and $TCRV\alpha/TCRV\beta$ chains due to the lack of large-scale data (74,83–86). Once more extensive data has become available, can we leverage machine learning to uncover the underlying structure of antigen-specific repertoires and the prediction rules of chain pairing? Uncovering these immunological prediction rules is crucial for the knowledge-based development of antibody and T-cell based immunotherapeutics.

**Resolving the sequence similarity architecture of immune repertoires**

The entirety of similarity relations among immune receptor sequences is called the similarity architecture of an immune repertoire. Thus, unlike immune repertoire diversity which is based on the frequency profiles of immune clones, sequence similarity architecture captures frequency-independent clonal sequence similarity relations. The similarity among immune receptors directly influences antigen recognition breadth: the more dissimilar receptors are, the larger is the antigen space covered. Given the genetic, cellular and clonal restrictions of immune repertoire diversity, the similarity architecture of antibody and T-cell repertoires has been a longstanding question and has only recently begun to be resolved. Understanding the sequence architecture of immune repertoires is, for example, crucial in the context of antibody therapeutics discovery for the conception of naïve antibody libraries and synthetic repertoires that recapitulate natural repertoires (87).

One powerful approach to interrogate and measure immune repertoire architecture is network analysis (Figure 1B, 2B) (87–93). Networks allow interrogation of sequence similarity and thereby add a complementary layer of information to repertoire diversity analysis. Clonal networks are built by defining each clone (nucleotide or amino acid sequence) as a node (Figure 1B). An edge between clones is drawn if they satisfy a certain similarity condition, which is predefined via a string distance (e.g., Levenshtein distance, LD) resulting in undirected Boolean networks (87–90,92,93). The default distance is usually 1 nucleotide or 1 amino acid difference but larger distances have also been explored (87). Thus, the construction of clonal networks requires the calculation of an all-by-all distance matrix. While the complete distance matrix can be computed on a single machine with repertoires of clone sizes <10,000, it becomes computationally expensive in terms of time and memory to calculate networks of clone sizes that exceed $10^5$ clones – which is the size of many repertoires in both mice and humans (3,5,6). Therefore, Miho et al. have developed a high-performance computing pipeline (*imNet*), which can compute distance matrices and construct corresponding large-scale repertoire networks (87). This method led to the biological insight that antibody repertoire networks are, in contrast to other systems (94,95), resistant to subsampling, which is of great importance for the network analysis of human repertoires where limited access to B-cell populations and lymphoid organs restricts complete biological sampling (17,79). Although networks of a few thousand nodes may be visualized using software suites such as igraph (96), networkx (97), gephi (98) and cytoscape (99), interpretation of the visual graphics is not informative for networks beyond a clonal size of $10^3$ (87). Furthermore, visualization of networks provides only marginal quantitation of the network similarity architecture thus limiting the quantitative understanding of immune repertoires. Graph properties and network analysis have been recently employed in order to quantify the network architecture of immune repertoires (87,93). Architecture analytics may be subdivided into properties that capture the repertoire at the global level (generally one coefficient per network), and those that describe the repertoire at the clonal and thus local level (one coefficient per clone per repertoire, vector of size equal to the clone size) (87).

Global coefficients are for example degree distribution, clustering coefficient, diameter and assortativity (87). The degree of a node is the number of its edges (i.e., the number of similar clones to a certain clone) and a repertoire's degree distribution quantifies the abundance of node degrees (i.e., clonal similarities) across clones of a repertoire. This degree distribution has been used to describe and classify networks by type, such as power-law (a few highly connected clones and many clones with few connections), which is reminiscent of antigen-driven clonal expansion, or exponential (more even degree distribution across clones, which cover an extensive sequence space) is more reflective of naïve repertoires (87). The degree distribution thus provides insights into the overall distribution of connectedness (clonal similarities) within a repertoire and its state of clonal sequence expansion. Local characterization allows for the interrogation and correlation of additional clonal related features, such as frequency and antigen specificity, within the immune repertoire architecture. Local parameters are for example: degree, authority, closeness, betweenness and PageRank (87). PageRank, for instance, measures the importance of the similarity between two CDR3 clones within the network. Detailed mathematical descriptions of available network parameters have been described elsewhere (87,100,101).

Complementary to networks, similarity indices have been devised that provide a continuous description of repertoire architecture by quantifying the similarity between all sequences of a repertoire (using distance metrics) on a scale ranging from 0 (zero similarity) to 100% (all sequences are 100% identical) (6,102). In addition to sequence similarity, the index by Strauli and Hernandez takes the frequency of each sequence into account thus normalizing sequencing similarity by the frequency of each of the pairwise compared sequences (102).

The assessment of repertoire architecture has only recently started to transition from the visual investigation of clusters of immune receptor sequences to the construction of large-scale networks and truly quantitative analysis of entire repertoires across similarity layers (> 1 amino acid/nucleotide differences). This advance enabled the discovery of fundamental properties of repertoire architecture such as reproducibility, robustness and redundancy (87). And although the biological interpretation of the mathematical characterization of immune repertoire networks is at an early stage, the universal use of network analysis in the deconvolution of complex systems (100,101) suggests a great potential in immune repertoire research. Many important questions remain: (i) How can network repertoire architecture be compared across individuals without condensing networks into network indices without potentially losing information? Thus, can discrete and continuous representation of repertoire architecture be merged into one comprehensive mathematical framework? (ii) Can the linking of networks across similarity layers serve to understand the dynamic and potential space of antigen-driven repertoire evolution (87)? (iii) Is the network structure that is observed on the antibody immunogenomic level, also maintained on the phenotypic and immunoproteomics level of serum antibodies (103–109)?

**Retracing the antigen-driven evolution of antibody repertoires**

Upon antigen challenge, B cells expand and hypermutate their antibody variable regions, thus forming a B-cell lineage that extends from the naïve unmutated B cells, to somatically hypermutated memory B cells (25), to terminally-differentiated plasma cells (11). Somatic hypermutation is unique to B cells and absent in T cells. Retracing antibody repertoire evolution enables insights into how vaccines (76) and pathogens shape the humoral immune response (110–112).

In order to infer the ancestral evolutionary relationships among individual B cells, lineage trees are constructed from the set of sequences belonging to a clonal lineage (Figure 1C, 2C). A clonal lineage is defined as the number of receptor sequences originating from the same recombination event. For building a lineage tree, a common preprocessing step is to group together all sequences with identical V, J gene and CDR3 length. Schramm et al. published a software for the ontogenetic analysis of antibody repertoires, which is designed to enable the automation of antibody repertoire lineage analysis. Importantly, it provides interfaces to phylogenetic inference programs such as BEAST and DNAML (113).

In antibody repertoire phylogenetics, there is no consensus as to which phylogenetic method is optimal for the inference of lineage evolution (17,114). Most of the current phylogenetic methods rely on assumptions that may be true for species evolution but might be invalid for antibody evolution. One prominent example is the assumption that each site mutates independently of the neighboring nucleotides, which is not the case in antibody evolution (114). Additionally, antibodies evolve on time scales that differ by several orders of magnitudes from those of species. These two factors likely decrease the accuracy of clade prediction (clade: set of descendent sequences that all share a common ancestor), thus potentially impacting antibody phylogenetic studies.

Several phylogenetic methods, such as Levenshtein distance (LD), neighbor-joining (NJ), maximum parsimony (MP), maximum-likelihood (ML), and Bayesian inference (BEAST) have been used for delineating the evolution of B-cell clonal lineages from antibody repertoire sequencing data (77,115–117). For general information regarding the methods, refer to the review by Yang and Rannala (118). Briefly, both LD and NJ are distance-based methods that rely upon an initial all-by-all distance matrix calculation and have been implemented in many computational platforms (Clustal, T-REX) and R packages (ape, phangorn) (119–122). Even in the event >$10^5$ sequences per sample, the distance matrix calculation in phylogenetics poses less of a problem than in network analysis since a sample's sequences are grouped by lineage members of identical V-J gene and CDR3 length thus reducing computational complexity. The relatively short computation-time of distance-based methods, renders them particularly useful for initial data exploration (118). Maximum parsimony attempts to explain the molecular evolution by non-parametrically selecting the shortest possible tree that explains the data (24). Maximum parsimony trees can be produced using several available tools (e.g., PAUP, TNT, PHYLIP, Rphylip) (123–126). Both ML and BEAST infer lineage evolution using probabilistic methods, which can incorporate biologically relevant parameters such as transition/transversion rate and nucleotide frequencies. A variety of ML tools have been developed (e.g., PhyML, RAxML, and MEGA) (127–129). While multiple phylogenetic tools utilizing Bayesian methods exist (130,131), this review focuses on BEAST given its recurrent use in antibody repertoire studies (113,117,132–134). BEAST traditionally employs a Markov chain Monte Carlo (MCMC) algorithm to explore the tree parameter space. This computationally expensive process limits the practical number of sequences per lineage tree to <$10^3$. Despite the extensive computational requirements (both in memory and in run time), BEAST has the advantage of producing time-resolved phylogenies and inferring somatic hypermutation rates (131,132). The BEAST framework shows, therefore, the highest scientific benefit when applied to experiments examining antibody evolution within the same host across multiple sampling time points (117), as inferred mutation rates and tree heights (duration of evolution) are reported in calendar time.

Yermanos et al. have compared five of the most common phylogenetics reconstruction methods for antibody repertoire analysis in terms of their absolute accuracy and their concordance in clade assignment using both experimental and simulated antibody sequence data (132). Correctly inferring the clades of a phylogenetic tree is crucial for describing the evolutionary relationship between clonally selected and expanded B-cells (i.e., memory B cells) that belong to a given lineage (i.e., derived from a naïve B cell). Phylogenetic trees inferred by the methods tested (LD, NJ, ML, MP, BEAST) resulted in different topologies as measured by both clade overlap (number of internal nodes sharing the same descendant sequences) and treescape metric (comparison of the placement of the most recent common ancestor of each pair of tips in two trees) (135). These results suggest caution in the interpretation and comparison of results from the phylogenetic reconstruction of antibody repertoire evolution (132).

The accurate reconstruction of antibody phylogenetic trees is tightly linked to the detailed understanding of the physical and temporal dynamics of somatic hypermutation along antigen-driven antibody sequence evolution. Mutation statistics can be inferred probabilistically to account for the fact that the likelihood of mutation is not uniformly distributed over the antibody VDJ region (45,46). For example, there is a preference to mutate particular DNA motifs called hotspots (length: 2–7bp) and concentrated in the CDRs over others (coldspots) (4,114,136,137). To uncover the sequence-based rules of somatic hypermutation targeting, Yaari et al. developed S5F, an antibody-specific mutation model. This model provides an estimation of the mutability and mutation preference for each nucleotide in the VDJ region of the heavy chain based on the four surrounding nucleotides (two on either side). The estimated profiles could explain almost half of the variance in observed mutation patterns and were highly conserved across individuals (114). Cui et al. have, in addition, reported two new models that add to the heavy chain S5F model: the light chain mouse RS5NF and the light chain human S5F L chain model (138). Additionally, Sheng et al. investigated the intrinsic mutation frequency and substitution bias of somatic hypermutations at the amino acid level by developing a method for generating gene-specific substitution profiles (139). This method revealed gene-specific substitution profiles that are unique to each human V-gene and also highly consistent between human individuals.

The existence of hotspot and coldspot mutation motifs violates the standard assumption of likelihood-based phylogenetics, which is that evolutionary changes at different nucleotide or codon-sites are statistically independent. Furthermore, since hotspot motifs are, by definition, more mutable than non-hotspot motifs, their frequency within the B-cell lineage may decrease over time as they are replaced with more stable motifs (140). In order to explicitly parameterize the effect of biased mutation within a phylogenetic substitution model, Hoehn et al. developed a model that can partially account for the effect of context-dependent mutability of hot- and coldspot motifs, and explicitly model descent from a known germline sequence (141). The resulting model showed a substantially better fit to three well-characterized lineages of HIV-neutralizing antibodies, thus being potentially useful for analyzing the temporal dynamics of antibody mutability in the context of chronic infection. In addition, Vieira et al. assessed the evidence for consistent changes in mutability during the evolution of B-cell lineages (133). Using Bayesian phylogenetic modeling, they showed that mutability losses were about 60% more frequent than gains (in both CDRs and FRs) in anti-HIV antibody sequences (133).

Although computational methods tailored to the phylogenetic analysis of antibody evolution are slowly beginning to surface, many important problems remain. (i) First approaches in coupling clonal expansion information to the inference of phylogenetic trees have been developed (142). Will these additional layers of information enable a better prediction of antibody evolution? (ii) There has been progress in comparing the differences of antibody repertoires in the context of phylogenetic trees using the UniFrac distance measure (143,144). Briefly, for a given pair of samples, UniFrac measures the total branch length that is unique to each sample. The comparison of tree *topologies*, however, remains a challenge. This is because each lineage tree is composed of a different number of sequences and there are thousands, if not more, of simultaneously evolving lineages within a single host. Although methods exists for the comparison of unlabeled phylogenetic trees by, for instance, means of their Laplacian spectra (145), their application and ability to extract meaningful biological conclusions have not yet been realized. (iii) It is unclear to what extent antibody evolution differs between different acute and chronic viral infections, or different antigens. Specifically, is it possible to relate antigen-driven convergence and affinity (6,49,110) to phylogenetic antigen-specific signatures (146)?

**Dissecting naïve and antigen-driven repertoire convergence**

Convergence (overlap) of immune repertoires describes the phenomenon of identical or similar immune receptor sequences shared by two or more individuals. Specifically, sequence convergence can either mean that (i) clones (public clonotypes, entire clonal sequence or clonotype cluster) or (ii) motifs (sequence substrings) are shared. Several researchers in the field have endeavored to quantify the extent of naïve and antigen-driven repertoire convergence using a large variety of computational approaches that quantify cross-individual sequence similarity (6,52,76,110,147–149) (Figure 1D, 2D). Repertoire convergence may be of substantial importance for the prediction and manipulation of adaptive immunity (6).

The simplest way to quantify sequence convergence is by clonotype overlap among pairwise samples expressed as a percentage normalized by the clonal size of either one or both of the samples compared (6,47,150). In case clonotypes are treated not as single sequences but clusters of sequences, clusters were defined as shared between samples if each sample contributed at least one sequence to the cluster (149). Overlap indices such as Morisita-Horn (151) add additional information to the measurement of clonal overlap by integrating the clonal frequency of compared clones (60,152,153). A parameterized version of the Morisita-Horn index, similarly to the Hill-diversity, may be used to weigh certain clonal abundance ranges differently (58). Rubelt and Bolen expanded on the idea of an overlap index by incorporating both binned sequence features (e.g., clone sequences, germline genes) and their frequency for measuring the impact of heritable factors on VDJ recombination and thymic selection. Their Repertoire Dissimilarity Index (RDI) is a non-parametric Euclidian-distance-based bootstrapped subsampling approach, which enables the quantification of the average variation between repertoires (49,154). Importantly, it accounts for variance in sequencing depth between samples. Another clone-based approach was developed by Emerson et al. who mined public TCRβ clonotypes in CMV-positive and CVM-negative individuals in order to predict their CMV-status. To this end, they identified CMV-associated clonotypes by using Fisher's exact test. Subsequently, these clonotypes were used within the context of a probabilistic classifier to predict an individual's CMV-status. The classifier used dimensionality reduction and feature selection in order to mitigate the influence of the variance of HLA types across individuals (since the distribution of TCRβ clones is HLA-dependent) (147).

Moving from the clonal to the subsequence-level, several groups compared the average distance between repertoires based on their entire sequence diversity (without predetermining feature bins). Specifically, Yokota et al. developed an algorithm for comparing the similarity of immune repertoires by projecting the high-dimensional inter-sequence relations, calculated from pairwise sequence alignments, onto a low-dimensional space (155). Such low-dimensional embedding of sequence similarity has the advantage of enabling the identification of those sequences that contribute most to inter-sample dis(similarity). As previously described, Strauli and Hernandez quantified sequence convergence between repertoires in response to influenza vaccination not only by incorporating genetic distance (Needleman-Wunsch algorithm) but also the frequency of each clonal sequence (102). Their approach relies on a statistical framework called functional data analysis (FDA), which is often used for gene-expression analysis. In their implementation, FDA models each sample as a continuous function over sampling time points and is thus suitable for the analysis of sequence convergence over a time-course experiment. The FDA framework has the advantage of accounting for uneven time-point sampling and measurement error, both of which are common characteristics of immune repertoire datasets (2,17). Bürckert et al. also employed a method borrowed from gene expression analysis (DESeq2) (156), to select for clusters of CDR3s, which are significantly overrepresented within different cohorts of immunized animals (157). These clusters exhibited convergent antigen-induced CDR3 signatures with stereotypic amino acid patterns seen in previously described tetanus toxoid and measles-specific CDR3 sequences.

Given the high-dimensional complexity of the immune repertoire sequence space, sequence-distance-based approaches might not suffice for covering the entire complexity of sequence convergence. A greater portion of the sequence space may be covered by sequence-based machine learning. Here the idea is that sequence signatures and motifs are shared between individuals belonging to a predefined class (e.g., different immune status). Sun et al. discriminated the TCRβ repertoire of mice immunized with and without ovalbumin with 80% accuracy by deconstructing it into overlapping amino acid k-mers (158). Cinelli et al. used a one-dimensional Bayesian classifier for the selection of features, which were subsequently used for support-vector machine analysis (159). As a third machine-learning alternative, Greiff et al. leveraged gapped-k-mers and support vector machines for the classification of public and private clones with 80% accuracy from antibody and TCR repertoires of human and mice. This study used overlapping k-mers to construct sequence prediction profiles, which highlight those convergent sequence regions that contribute most to the identity of a class (public/private clones but also e.g., also different immune states and antigen specificities) (56). Beyond k-mers, several groups have exploited the addition of additional information such as physico-chemical properties (Atchley and Kidera factors), in order to provide more extensive information to machine-learning algorithms (160–164). Finally, a machine-learning independent approach using local-search graph theory for the detection of disease-associated k-mers was recently published by Apeltsin et al. (165).

One of the longest-standing challenges in immunology is whether it is possible to predict antigen specificity from the sequence of the immune receptor (2,15,166–168). Sequence-dependent prediction implies that immune receptor sequences specific to one antigen share exclusive sequence signatures (motifs) or have higher intra-class than inter-class similarity (class = antigen). Two investigations in the direction of sequenced-based specificity prediction have recently been undertaken using sequence similarity (sequence distance) approaches (148,169). In one example, Dash et al. developed a distance measure, called TCRDist, which is guided by structural information on pMHC binding (148). Two TCRs sequences were compared by computing a similarity-weighted Hamming distance between CDR sequences, including an additional loop between CDR2 and CDR3. TCRDist was used to detect clusters of highly similar, antigen-specific groups of TCRs that were shared across different mouse or human samples. In order to predict the antigen-specificity of a TCR, it was assigned to the cluster to which it had the highest similarity (as based on the TCRDist), resulting in highly accurate prediction (148). Using a similar approach, Glanville et al. developed GLIPH, a tool that identifies TCR specificity groups using a three-step procedure: (i) determining of shared motifs and global similarity, (ii) clustering based on local and global relationships between TCRs and (iii) analyzing the enrichment for common V-gene, CDR3 lengths, clonal expansion, shared HLA alleles in recipients, motif significance, and cluster size. This approach yielded also highly accurate prediction of antigen-specific TCRs and led to the design of synthetic TCRs (not existing in biological data) that retained antigen specificity (169).

One of the biggest bottlenecks of learning the underlying principles of antigen-driven repertoire convergence is the scarcity of antigen-specific sequence data. This is not only a problem for machine learning but also network-based approaches, where one wishes to map antigen-specific information onto generated networks (87,93). To address this issue for T cells, Shugay et al. (VDJDB) and Tickotsky et al. (McPAS-TCR) have built dedicated and curated databases. VDJDB gathers >10,000 TCR sequences from different species associated with their epitope (>200) and MHC context (170). McPAS-TCR contains more than 5000 pathogen-associated TCRs from humans and mice (171). For antibodies, Martin et al. have conceived Abysis, which encompasses >5000 sequences of known function (from literature) from many species (>15) along with, where available, PDB 3D-structure information (172).

Significant progress in the understanding of antigen-associated signatures has been made. However, several long-standing questions remain to be answered: (i) The emergence of antigen-driven convergence and phylogenetic evolution are inherently linked. Is it feasible to model both phenomena in a unified computational environment similarly to recent efforts in coupling phylogenetics with the understanding of somatic hypermutation patterns (133,141)? (ii) Can recently developed models for the inference of VDJ recombination patterns and selection factors be applied to the analysis of antigen-associated sequence signatures (4,54)? (iii) Do more advanced sequence-based machine learning techniques such as deep neural networks, capable of capturing long-range sequence interactions (out of reach for k-mer-based approaches), improve modeling of the epitope and paratope space (82,173–178)?

**Conclusion**

The toolbox of computational immunology for the study of immune repertoires has reached an impressive richness leading to remarkable insights into B-and T-cell development (6,179), disease and infection profiling (76,77), propelling forward the fields of immunodiagnostics and immunotherapeutics (63,111,180). Here, we have discussed computational, mathematical and statistical methods in the light of underlying assumptions and limitations. Indeed, although considerably matured over the last few years, the field still faces several important and scientifically interesting problems: (i) There exist only few platforms to benchmark computational tools, thus hindering the standardization of methodologies. Recently, a consortium of scientists working in AIRR-seq has convened to establish and implement consensus protocols and simulation frameworks (http://airr.irmacs.sfu.ca/) (2,17,33,34,42). (ii) With the exponential increase of both bulk and single cell data (83,181), the scalability of computational tools is becoming progressively important. Although advances in this regard have been made in sequence annotation, clonotype clustering and network construction (62,87,182,183), further efforts especially in the field of phylogenetics are necessary to infer the evolution of large-scale antibody repertoires (132). (iii) Although there exist many approaches, which capture part of the immune repertoire complexity, a computational approach for the synthesis of many dimensions of the repertoire space at once is missing thus hindering a high-dimensional understanding of the adaptive immune response. (iv) Very few attempts exist yet, which aim to link immune receptor and transcriptomics data (184,185). Recently, computational tools have been developed that can extract immune-receptor sequences from bulk and single-cell transcriptomic data (182,185–189). Linking immune repertoire and transcriptome may provide a deeper understanding of how antibody and T-cell specificity is regulated on the genetic level with profound implications for synthetic immunology (190–192). (v) Many methods capture a static space of repertoires and few methods create *predictive* quantitative knowledge. Increasing the predictive performance of computational methods will help in the antibody discovery from display libraries and immunizations, and the design of vaccines and immunodiagnostics (15,19,193–195).

**Conflict of interest statement**

The authors declare no conflict of interest.

**Funding Statement**

This work was funded by the Swiss National Science Foundation (Project #: 31003A_170110, to STR), SystemsX.ch – AntibodyX RTD project (to STR); European Research Council Starting Grant (Project #: 679403 to STR). The professorship of STR is made possible by the generous endowment of the S. Leslie Misrock Foundation. We are grateful to ETH Foundation for the Pioneer Fellowship to Enkelejda Miho.

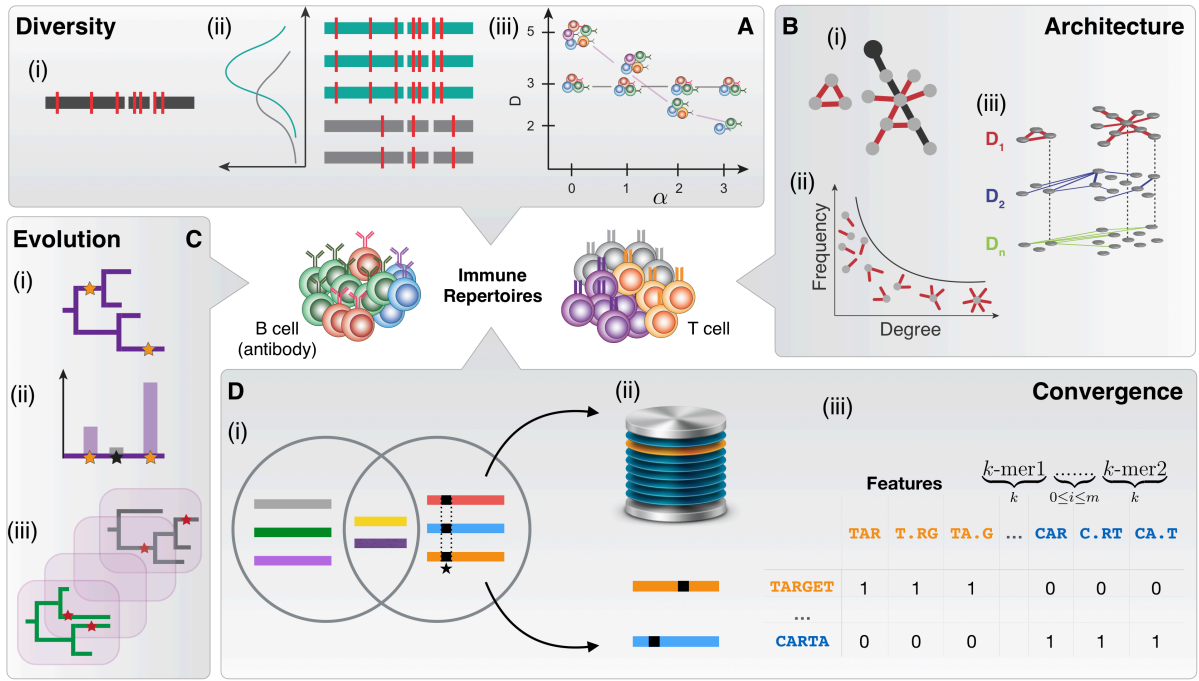

**Figure 1 The immune repertoire space is defined by diversity, architecture, evolution and convergence.**

**(A)** Diversity measurements are based on (i) the accurate annotation of V (D) J segments using deterministic and probabilistic approaches with population-level or individualized germline gene reference databases. (ii) Probabilistic and hidden Markov models allow inference of recombination statistics. (iii) Measurement of clonotype diversity using diversity profiles.

**(B)** Analysis of repertoire architecture relies predominantly on (i) clonal networks that are constructed by connecting nucleotide or amino acid sequence-nodes by similarity-edges. The sequence-similarity between clones is defined via a string distance (e.g., Levenshtein distance, LD) resulting in undirected Boolean networks for a given threshold (nucleotides/amino acids). An example of the global characterization of the network is the diameter, shown from the edges in black. An example of the local parameters of the network is the degree (n=1) related to the individual clonal node in black. (ii) Degree distribution is a global characteristic of immune repertoire networks, which can be used for analyzing clonal expansion. (iii) Several similarity layers decompose the immune repertoire along its similarity layers. Layer D1 captures clonal nodes similar by edit distance 1 (1 nt./a.a. different) , D2 of distance 2 and so forth.

**(C)** Assessing evolution of antibody lineages. (i) Reconstruction of phylogenetic trees. Stars indicate somatic hypermutation. (ii) Probabilistic methods for the inference of mutation statistics in antibody lineage evolution. (iii) Simulation of antibody repertoire evolution for benchmarking antibody-tailored phylogenetic inference algorithms.

**(D)** Naïve and antigen-driven cross-individual sequence similarity and convergence in immune repertoires. (i) The Venn diagram shows sequences shared in the two repertoires (circles). Signature-like sequence-features are highlighted by black squares. (ii) Database of convergent immune repertoire sequences. (iii) K-mer sequence decomposition and classification of immune repertoire sequences.

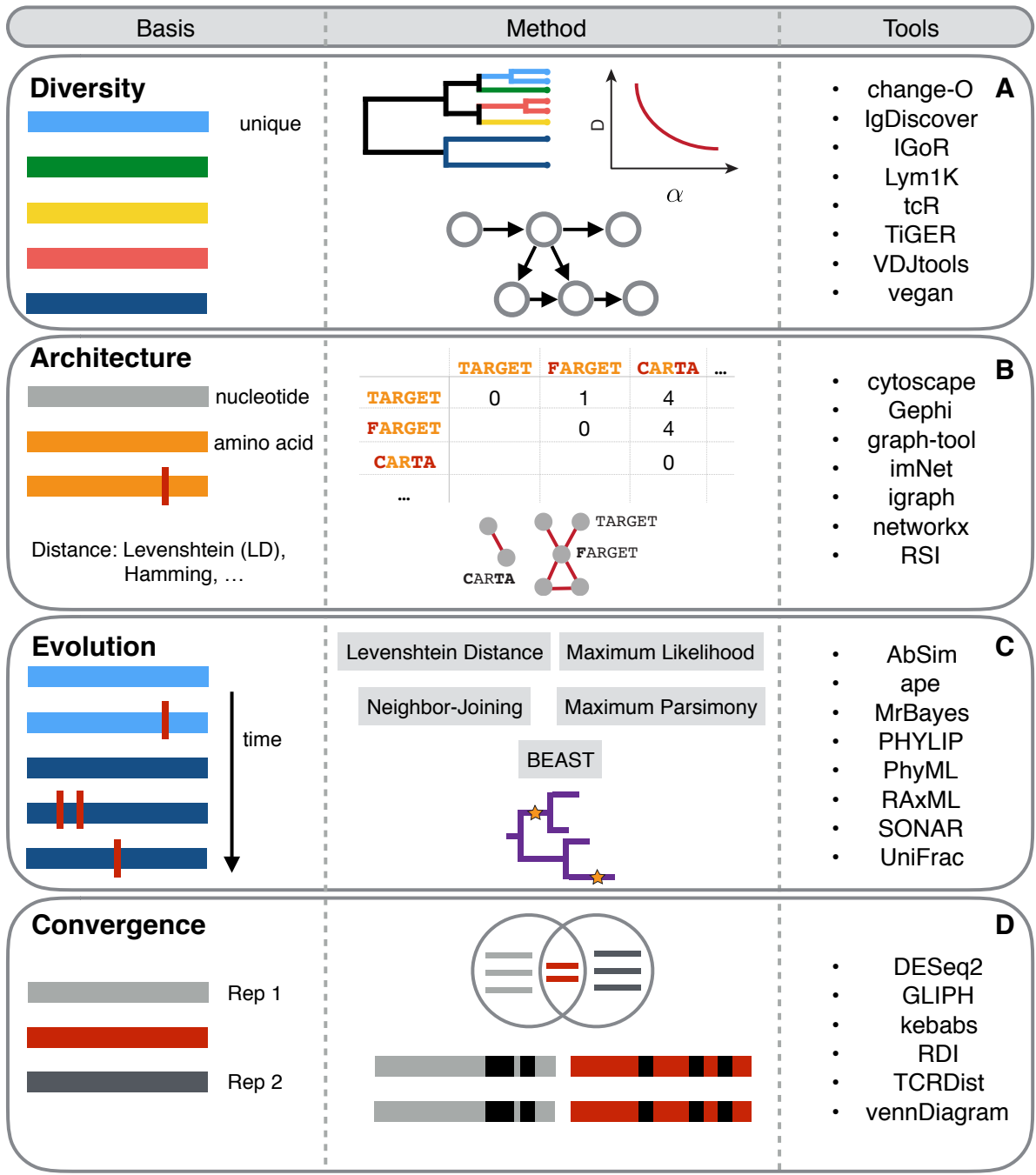

**Figure 2 An overview of selected computational tools used in immune repertoire analyses.** Each colored bar in the *Basis* column represents a unique antibody or TCR sequence. Red bars across represent sequence differences or somatic hypermutation. The *Method* column describes the general concept of the computational methods and how these are applied to immune repertoires. The *Tools* column highlights exemplary key resources for performing computational analysis in the respective analytical sections (rows, **A–D**).